\documentclass[aps,pra,floatfix,twocolumn,showpacs]{revtex4}

\usepackage{graphicx}
\usepackage{amssymb}
\usepackage{amsmath}

\newcommand {\ket}[1] {|#1 \rangle}

\begin{document}

\title{Quantum Computing and Quantum Simulation with Group-II atoms}

\author{Andrew J. Daley}
\affiliation{Department of Physics and Astronomy, University of Pittsburgh, Pittsburgh, Pennsylvania 15260, USA}

\begin{abstract}
Recent experimental progress in controlling neutral group-II atoms for optical clocks, and in the production of degenerate gases with group-II atoms has given rise to novel opportunities to address challenges in quantum computing and quantum simulation. In these systems, it is possible to encode qubits in nuclear spin states, which are decoupled from the electronic state in the $^1$S$_0$ ground state and the long-lived $^3$P$_0$ metastable state on the clock transition. This leads to quantum computing scenarios where qubits are stored in long lived nuclear spin states, while electronic states can be accessed independently, for cooling of the atoms, as well as manipulation and readout of the qubits. The high nuclear spin in some fermionic isotopes also offers opportunities for the encoding of multiple qubits on a single atom, as well as providing an opportunity for studying many-body physics in systems with a high spin symmetry. Here we review recent experimental and theoretical progress in these areas, and summarise the advantages and challenges for quantum computing and quantum simulation with group-II atoms. 
\end{abstract}

\date{June 25, 2011}
\maketitle

\section{Introduction}

Group-II and group-II-like species have recently become extremely interesting candidates to address challenges in quantum computing and quantum simulation with neutral atoms. This is motivated by rapid experimental progress, especially the control over these species developed in the context of  optical clocks \cite{AEatom1,zeemanshift1,ludlow08,barber08,clockreview}, and through the recent production of degenerate Bose and Fermi gases with Ytterbium \cite{takasu03,takasu07,Takahashi,yblattice1,yblattice2}, Calcium \cite{Kraft2009} and Strontium \cite{Stellmer2009,deEscobar2009,DeSalvo2010,Tey2010dgb}. In addition to the traditional advantages of neutral atoms in producing large arrays of qubits (e.g., loading a quantum register in an optical lattice), these species offer the opportunity to encode qubits in nuclear spin states that are robust against decoherence caused by fluctuating magnetic fields, and they exhibit an electronic transition to metastable levels that can be decoupled from the nuclear spin state, and used for readout and manipulation of the qubit. In this context, many techniques that were proposed for group-I atoms can be freed from important technical restrictions, creating new means to address challenges in slowing decoherence, producing high-fidelity quantum gates, and individually addressing specific qubits. Here we review the recent theoretical and experimental progress in quantum computing with group-II atoms, summarising the advantages they give over schemes with group-I atoms, as well as briefly discussing the progress in quantum simulation with these species.

\begin{figure}[tb]
\includegraphics[width=8.5cm]{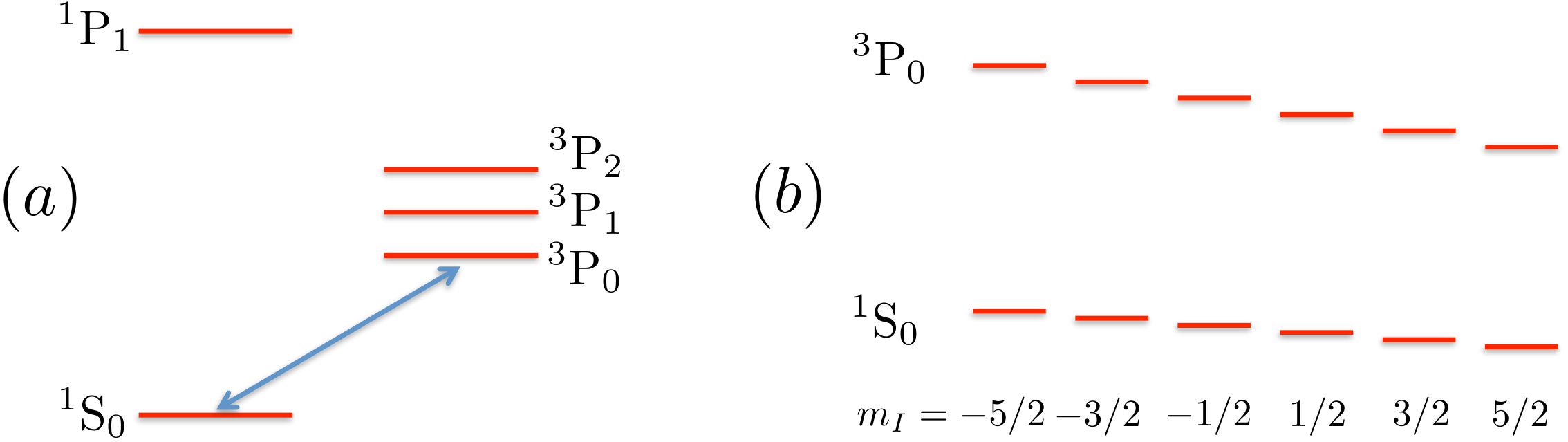}
\caption{Schematic level diagram for the electronic structure of group-II atoms, and related species such as Yb. (a) These atoms exhibit a singlet-triplet transition with long-lived metastable levels, including a clock transition between the ground $^1$S$_0$ manifold and the excited $^3$P$_0$ manifold. (b) While bosonic isotopes of group-II atoms have even-even nuclei and thus zero nuclear spin, fermionic isotopes have non-zero nuclear spin. For example, $^{173}$Yb has nuclear spin $I=5/2$, leading to different sublevels of the $^1$S$_0$ and $^3$P$_0$ manifolds, essentially corresponding to states that differ only in their nuclear spin (see text for more details).} \label{fig:levelstructure}
\end{figure}

The main advantages of group-II (alkaline-earth-metal) atoms over group-I (alkali-metal) atoms arise from the singlet-triplet metastable transition they exhibit (see Fig.~\ref{fig:levelstructure}), in which the $^1$S$_0$ -- $^3$P$_0$ plays the role of the clock transition in optical clock experiments. Typical lifetimes of the $^3$P$_0$ level are $\sim30$s (for $^{87}$Sr), and lifetimes in the other metastable $^3$P$_2$ level are longer still. While bosonic isotopes of group-II elements have zero nuclear spin as they have even-even nuclei, the fermionic isotopes have non-zero nuclear spin, and this nuclear spin can be decoupled from the electronic state on the clock transition \cite{childress, reichenbach07, yi08, dereviankoaddressing, aeshort, aelong, Shibata, Gorshkov2009} where the electronic angular momentum is zero. This has motivated a series of quantum computing schemes for group-II atoms  \cite{aeshort,Gorshkov2009,dereviankoaddressing,Shibata,aelong} in which the nuclear spin states are used to store qubits \cite{childress,hayes07,dereviankoaddressing}, due to their relative insensitivity to decoherence from magnetic field fluctuations, and these qubits can be manipulated or read out via coupling to the electronic state on the clock transition. For example, the metastable states can be used  to produce state-selective traps for collisional gate schemes \cite{aeshort, aelong}, to cool of the motion of the atoms in a manner that does not affect the stored qubits \cite{reichenbach07}, or for individual addressing of qubits using a magnetic gradient field that shifts the energy of the $^3$P$_2$ levels \cite{dereviankoaddressing, aeshort,Shibata}.

While group-II atoms could be used in place of group-I atoms in many quantum computing proposals, we will focus here primarily on schemes based on a quantum register of group-II atoms trapped with one atom per site in an optical lattice. This takes advantage of the possibility to form large quantum registers with neutral atoms, and is a good context in which to demonstrate the primary new features of group-II atoms. In addition, this is strongly motivated by recent experiments in which fermionic Yb has been loaded into optical lattice potentials in band insulating states, corresponding to arrays in which with a high probability each site is occupied by a single atom \cite{yblattice2}. Progress is also being made in this direction for Sr, with the recent production of degenerate Fermi gases of $^{87}$Sr \cite{DeSalvo2010,Tey2010dgb}. Group-II atoms in optical lattices are also a promising route to analogue quantum simulation, especially because of the possibility to construct models with high symmetry due to interparticle interactions being independent of the nuclear spin \cite{Cazalilla,Gorshkov}. At the same time the unique properties of group-II atoms offer new techniques for manipulation of atoms in the context of quantum simulation, ranging from new methods for production of artificial magnetic fields \cite{Gerbier2010} to opportunities for engineering dissipative processes for preparation of interesting many-body states \cite{DiehlYi2010}.

The rest of this review is organised as follows: In Sec.~\ref{sec:elements} we outline the different elements of potential quantum computing schemes with group-II atoms, discussing the differences and advantages to schemes with other neutral atoms, and outlining the current state of the art in experiments. In Sec.~\ref{sec:completeanalysis} we summarise these ideas by discussing possible complete schemes for quantum computing with group-II atoms, and providing an analysis of likely imperfections in experiments. In Sec.~\ref{sec:simulation} we discuss briefly the use of group-II atoms in analogue quantum simulation, outlining their advantages and special characteristics for this purpose. In Sec.~\ref{sec:summary} we provide a summary and outlook.

\section{Elements of quantum computing schemes for group-II atoms}
\label{sec:elements}
In this section we survey the unique advantages that group-II atoms provide for quantum computing schemes, discussing the theoretical proposals making use of these advantages, and also the current state of the art in experiments. We begin by discussing how nuclear spin states and electronic states of group-II atoms can be independently controlled, and the experimental possibilities that this provides. We then discuss specific elements of quantum computing schemes with group-II atoms, namely preparation of a quantum register, implementation of gate operations, and addressing of qubits for manipulation and readout. We also discuss the possibilites that arise for encoding multiple qubits on a single atom. 

\subsection{Independent manipulation of nuclear spin and electronic states}
\label{subsec:control}

\begin{figure}[tb]
\includegraphics[width=4.5cm]{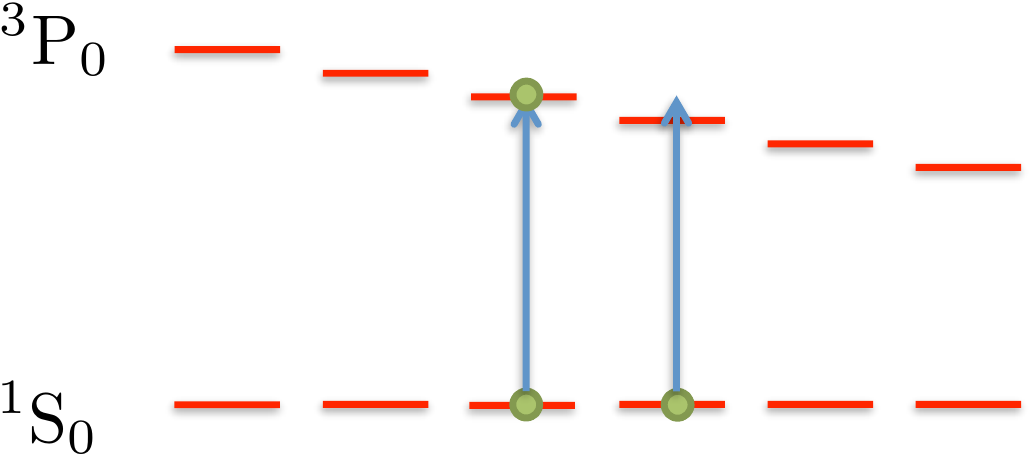}
\caption{Schematic diagram showing how nuclear spin states can be independently controlled. As performed, e.g., in Ref.~\cite{ludlow08}, atoms can be excited from the $^1$S$_0$ manifold to the $^3$P$_0$ manifold nuclear-spin-dependently by application of an external magnetic field. The differential Zeeman shift in the two levels then makes all but one transition off-resonant. Here, coupling is performed with $\pi-$polarised light. As the $^1$S$_0$-$^3$P$_0$ transition is weakly allowed only due to hyperfine mixing of higher-lying $P$ levels in the $^3$P$_0$ state, circularly polarised transitions (changing $m_F$) are also possible. (see text for more details).} \label{fig:manipulation}
\end{figure}

\subsubsection{Level structure and control}
The key to most quantum computing proposals with fermionic group-II atoms is the possibility to independently control the nuclear spin and electronic degrees of freedom \footnote{Bosonic isotopes of group-II elements have zero nuclear spin, as they have even-even nuclei.}. Because the electronic angular momentum is zero on the clock transition (i.e., in the $^1$S$_0$ and $^3$P$_0$ manifolds), these levels split into $2I+1$ sublevels for a species with nuclear spin $I$, as depicted in Fig.~\ref{fig:levelstructure}b. The species currently available in the laboratory exhibit a range of nuclear spin values, from $I=1/2$ ($^{171}$Yb) and $I=5/2$ ($^{173}$Yb) to $I=9/2$ ($^{87}$Sr). To a large extent, these nuclear spins decouple from the electronic state on the clock transition, especially in the presence of a large magnetic field. In the ground $^1$S$_0$ level, the sublevels are essentially states of fixed nuclear spin projection $m_I$ for all field strengths. It is important to note that the coupling matrix elements between $^1$S$_0$ and $^3$P$_0$ for fermionic species are essentially generated by hyperfine interactions, which lead to admixtures of higher lying $P$ states in the $^3$P$_0$ level \cite{hoyt05,porsev04}. The sublevels of $^3$P$_0$ are then perhaps better seen as states of fixed total angular momentum projection, $m_F$, even though they are approximately states of fixed nuclear spin projection $m_I$. As a result, optical coupling between the $^1$S$_0$ and $^3$P$_0$ can be generated with either circularly polarized or $\pi-$polarised light, with the resulting change in angular momentum projection, $\Delta m_F=\pm 1, 0$. In addition, a differential Zeeman shift between the $^1$S$_0$ and $^3$P$_0$ manifolds (200 Hz/G for $^{87}$Sr) means that in a moderate magnetic field \cite{boyd07}, atoms may be transferred spin-dependently between these two manifolds (see Fig.~\ref{fig:manipulation}). This type of control over the $m_F$ state and spin-dependent transfer was demonstrated in optical clock experiments in Boulder \cite{ludlow08}, and is the building block required to allow arbitrary manipulation of the combination of electronic and nuclear spin states on a single atom. Note that the timescales for such mechanisms is limited by the need to resolve transitions for a single qubit state. These operations are thus faster in stronger magnetic fields.

\subsubsection{Separate use of the electronic and nuclear spin states}
\label{subsubsec:separateuse}
This decoupling and independent manipulation of the electronic and nuclear spin states motivates several schemes in which qubits are stored in the nuclear spin states (with correspondingly long coherence times), and the electronic state is used for manipulation of the qubit. For example, such schemes have been proposed for use in cooling atomic motion \cite{reichenbach07}, and the production of spin-dependent lattices for quantum gates \cite{aeshort,aelong}. The $^3$P$_2$ level can also be used for individual addressing of qubits \cite{dereviankoaddressing,aeshort,Shibata,aelong} as described in Sec.~\ref{subsec:addressing} below.

\begin{figure}[tb]
\includegraphics[width=6.5cm]{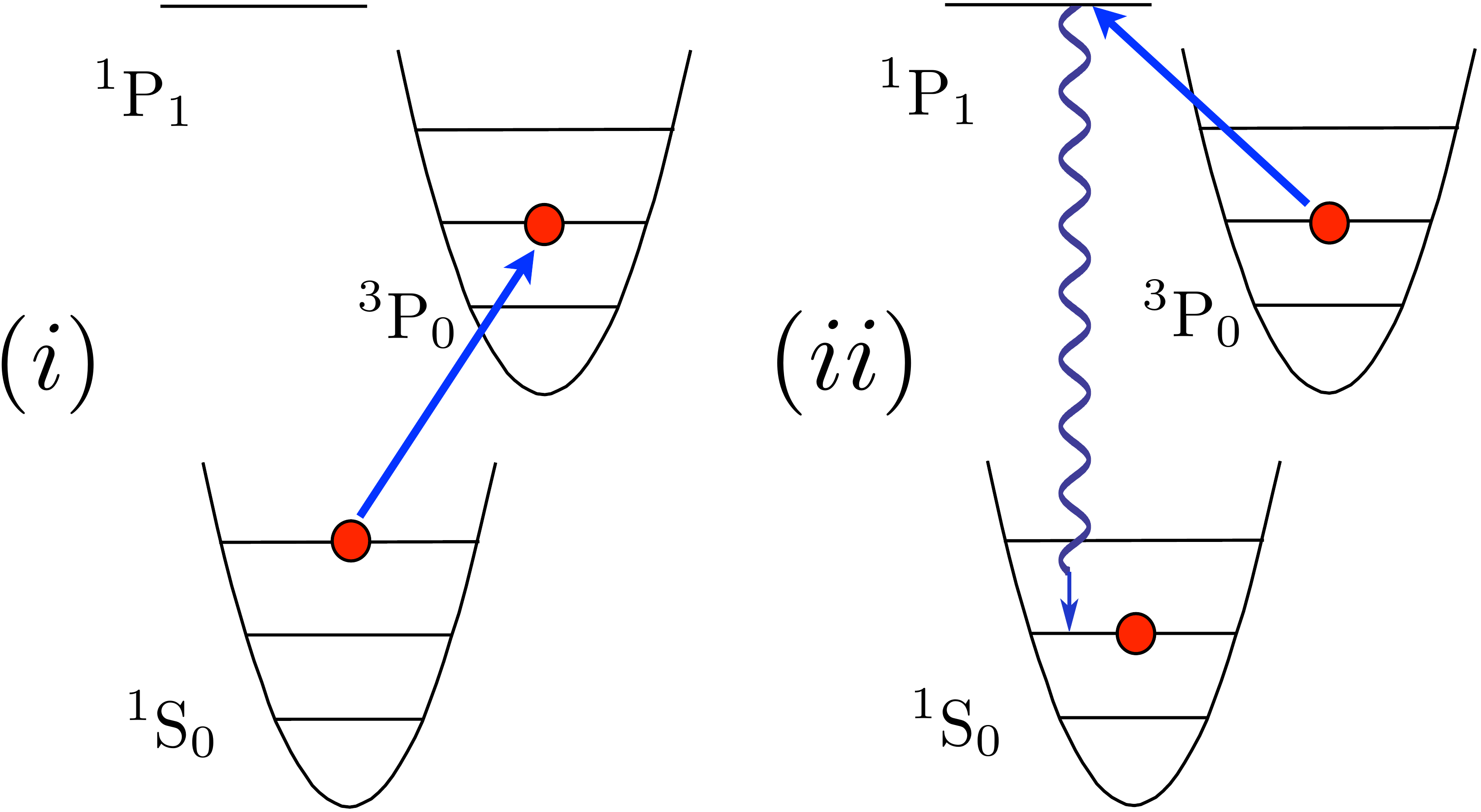}
\caption{Schematic diagram of the two key steps in the resolved sideband cooling scheme proposed by Reichenbach and Deutsch \cite{reichenbach07}. (i) Atoms are coupled from the $^1$S$_0$ manifold to the $^3$P$_0$ manifold on the red sideband for the motional levels, which can be resolved because of the narrow linewidth in $^3$P$_0$. (ii) The metastable triplet state is then quenched by coupling to the $^1$P$_1$ manifold. (Nuclear spin degrees of freedom are omitted here, see Fig.~\ref{cooling2} and text for more details).} \label{cooling1}
\end{figure}

\begin{figure}[tb]
\includegraphics[width=6.5cm]{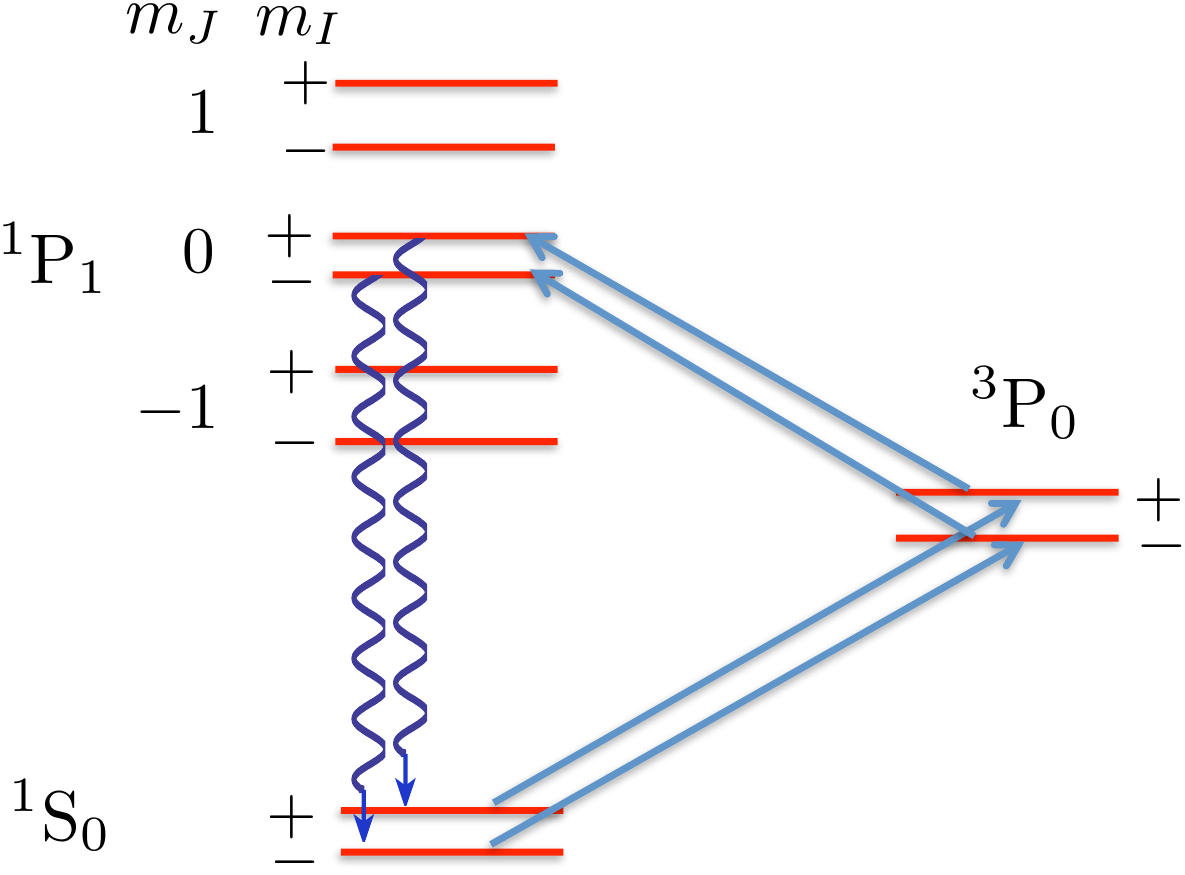}
\caption{Level scheme for the resolved sideband cooling scheme proposed by Reichenbach and Deutsch \cite{reichenbach07}. Vibrational levels are not shown (see Fig.~\ref{cooling1}. In order to preserve nuclear spin coherence, atoms are coupled from the $^1$S$_0$ manifold to the $^3$P$_0$ manifold with $\pi$ transitions. A large magnetic field is applied so that in the $^1$P$_1$ manifold the states are essentially product states of electronic angular momentum and nuclear spin, which allows the quench of $^3$P$_0$ to return atoms to $^1$S$_0$ without destroying qubit states encoded on the nuclear spin (see text for more details).} \label{cooling2}
\end{figure}

\emph{Non-destructive qubit cooling -} A good example of how the decoupling of electronic and nuclear spin states can be used is provided by the cooling scheme proposed by Reichenbach and Deutsch \cite{reichenbach07}, in which they demonstrate that the motion of trapped atoms could be cooled whilst preserving the state of nuclear spin qubits. Such a process is not possible for group-I-like atoms, where spontaneous emission events erase qubit states. This scheme is based on resolved sideband cooling \cite{resolvedsidebandcooling}, as depicted in Fig.~\ref{cooling1}. An atom beginning in the $^1$S$_0$ level is excited to the $^3$P$_0$ level, with coupling on a resolved red sideband for the motion (i.e., it is transferred to a lower vibrational level in the trap). The trap is assumed to be identical for the two levels, which can be achieved, e.g., by using a \emph{magic wavelength lattice} \cite{AEatom1, zeemanshift1}, where the polarisability of $^1$S$_0$ and $^3$P$_0$ are equal for the wavelength of light creating the lattice. The atom is then returned to the $^1$S$_0$ level via an incoherent process, in this case spontaneous decay (see Fig.~\ref{cooling1}). In order to preserve the nuclear spin state during this process, both the coherent excitation and spontaneous decay must preserve the nuclear spin. As indicated above, the nuclear spin can be preserved in the coherent excitation by using $\pi$-polarised light. It is also important to ensure that all coherent superpositions are faithfully transferred to the $^3$P$_0$ level even when the transitions are shifted for different $m_I$ due to an applied external magnetic field, by either using multiple pulses at the corresponding frequencies, or adiabatic transfer techniques. In the spontaneous decay step, it is not sufficient to 
allow an atom to decay from $^3$P$_0$ to $^1$S$_0$. This process would not only be very slow, but because the $^3$P$_0$-$^1$S$_0$ is weakly allowed because of hyperfine mixing (see above), spin-changing transitions are allowed. Instead, the $^3$P$_0$ level is quenched by pumping atoms to the $^1$P$_1$ level, as depicted in Fig.~\ref{cooling2}, which decays with high probability to the $^1$S$_0$ level. By applying a large magnetic field, the states in the $^1$P$_1$ level can be split essentially into product states of fixed electron angular momentum projection $m_J$ and nuclear spin projection $m_I$. In this limit, atoms can be quenched to the $m_J=0$ level, and will decay to the $^1$S$_0$ with no decoherence for the nuclear spin state.

Given vibrational frequencies of the order of $\sim 100$ kHz, and the small linewidth of the clock transition (of the order of $\mu$Hz), extremely small probabilities of vibrational excitations ($\lesssim 10^{-15}$) could theoretically be achieved using this scheme. In order to avoid decoherence of the nuclear spin state, it is important to achieve high enough magnetic fields that electronic angular momentum and nuclear spin decouple in the $^1$P$_1$ manifold. To reach a fidelity of $99\%$ per cooling cycle, a field of $1$ T would be required for $^{171}$Yb, though only $1$ mT for $^{87}$Sr \cite{reichenbach07}.

\emph{Electronic state readout -} Similar ideas can also be used for reading out the electronic state without generating significant decoherence for the nuclear spin state. If a large magnetic field is applied so that the electronic angular momentum and nuclear spin decouple in the $^1$P$_1$ manifold, then the $^1$S$_0$-$^1$P$_1$ transition can be driven close to resonance. Photons will then be scattered if the atom is present in the $^1$S$_0$ level, whilst the nuclear spin state will be preserved. An alternative that will also work in weaker magnetic fields is driving the same transition far off resonance compared with the level splitting in $^1$P$_1$ \cite{childress}. In this limit, paths for photon scattering events that result in a change of the nuclear spin state destructively interfere, and the qubit state would again be preserved. This second scheme is analysed in detail by Gorshkov et al. \cite{Gorshkov2009} in the context of producing a quantum register with group-II atoms.

\begin{figure}[tb]
\includegraphics[width=6.5cm]{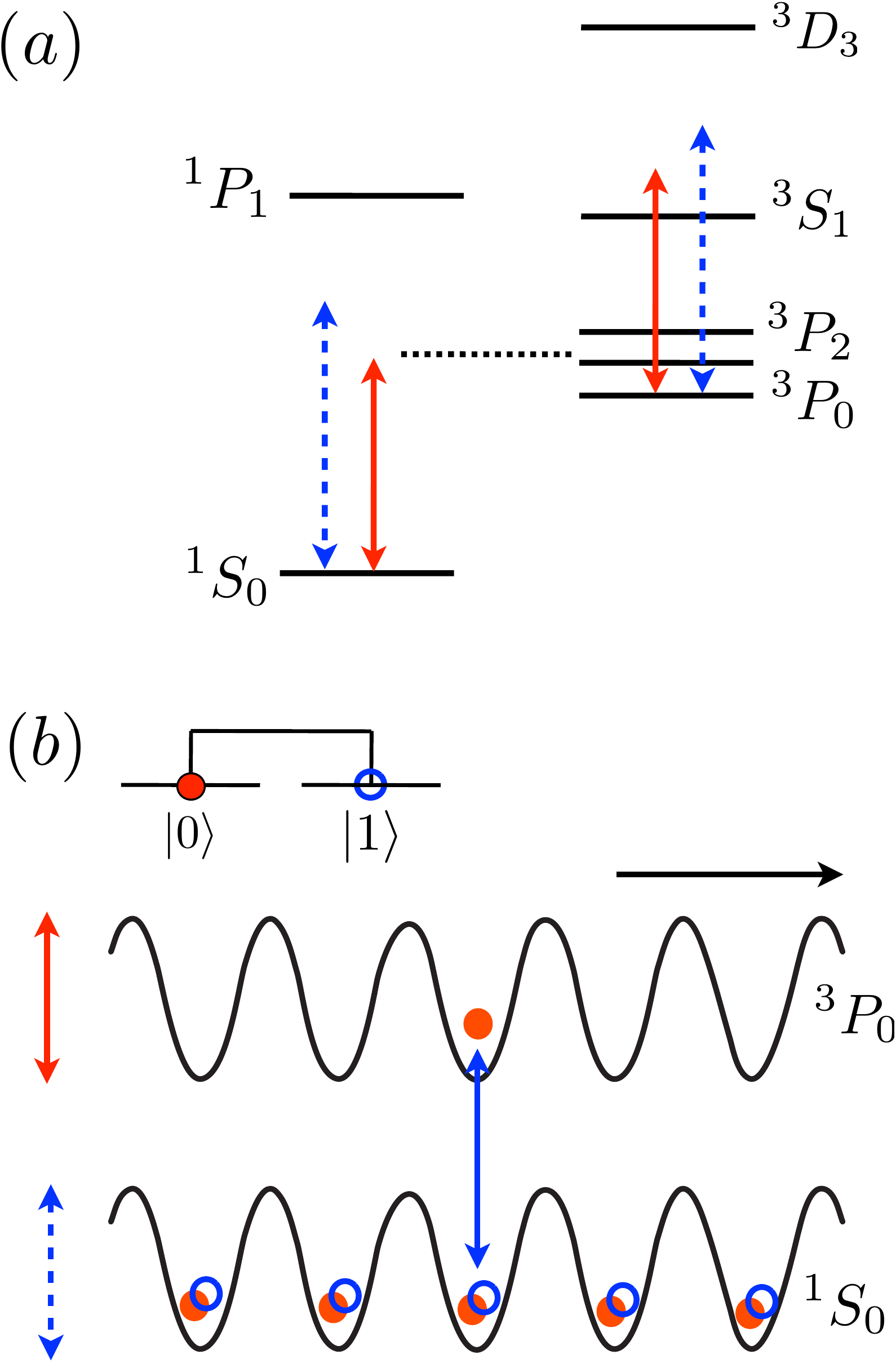}
\caption{Making electronic state-dependent lattices. (a) By choosing appropriate frequencies of light it is possible to create lattices from lasers that provide an AC-Stark shift only for the $^1$S$_0$ manifold and not for the $^3$P$_0$ manifold (dashed lines), or only for the $^3$P$_0$ manifold and not for the $^1$S$_0$ manifold (solid lines). (b) These lattices can be used, e.g., for storage of qubits (encoded on the nuclear spin) in the $^1$S$_0$ lattice, and transport of atoms to distant sites for gate operations in the $^3$P$_0$ lattice, as discussed in Ref.~\cite{aeshort}.} \label{spindeplattice}
\end{figure}

\emph{Spin-dependent lattices for qubits -} Another important use for the decoupling of the electronic and nuclear spin states is the creation of state-dependent lattices. For example, the $^1$S$_0$ ground state and $^3$P$_0$ metastable state belong to different transition families and are separated by optical frequencies, resulting in them exhibiting considerably different AC polarisabilities as a function of the wavelength of applied light. As a result, it is possible to find two wavelengths where optical traps can be generated via the AC-Stark shift for each of these states completely independently of the other, as depicted in Fig.~\ref{spindeplattice}. In Ref.~\cite{aeshort} it is shown, e.g., that for $^{87}$Sr, the polarisability of $^3$P$_0$ is zero at 627 nm, because of cancelling shifts of different signs from more highly excited triplet levels, whilst the polarisability of $^1$S$_0$ is $\sim 430$ a.u.. Light at this wavelength can be used to form a deep optical lattice for $^1$S$_0$ which will not affect the $^3$P$_0$ states, and can be used, e.g., as a \textit{storage} lattice for qubits. Similarly, for $^{87}$Sr, the polarisability of $^1$S$_0$ at 689.2 nm is zero, whereas the polarisability $^3$P$_0$ is $\sim 1550$ a.u.. This is primarily because of the near-resonant coupling of $^1$S$_0$ to $^3$P$_1$, which despite being near-resonant does not exhibit large spontaneous emission rates as the linewidth of $^3$P$_1$ is very narrow. This lattice, for example, be used for \textit{transport} of atoms, without affecting atoms stored in the $^1$S$_0$ state. In order to make overlapping lattices, the potentials should be given the same spatial period by using angled beams in the case of the 627 nm light. As noted below, gate operations can then be performed between qubits stored in distant sites by transferring atoms state-selectively into the transport lattice, and moving them to the appropriate site. This is motivated by the use of spin-dependent lattices for group-I atoms \cite{spindep1,spindep2}, but frees this scheme from an important technical restriction: In group-I atoms, spin-dependent lattices are formed by tuning lattice lasers between fine-structure states, which can lead to large heating and decoherence from spontaneous emissions. Here, the lattices for different spins are made completely independent by selection of the appropriate wavelengths, without the same complications in terms of heating. In addition, these schemes can be applied unmodified in any number of dimensions. 

\begin{figure}[tb]
\includegraphics[width=6.5cm]{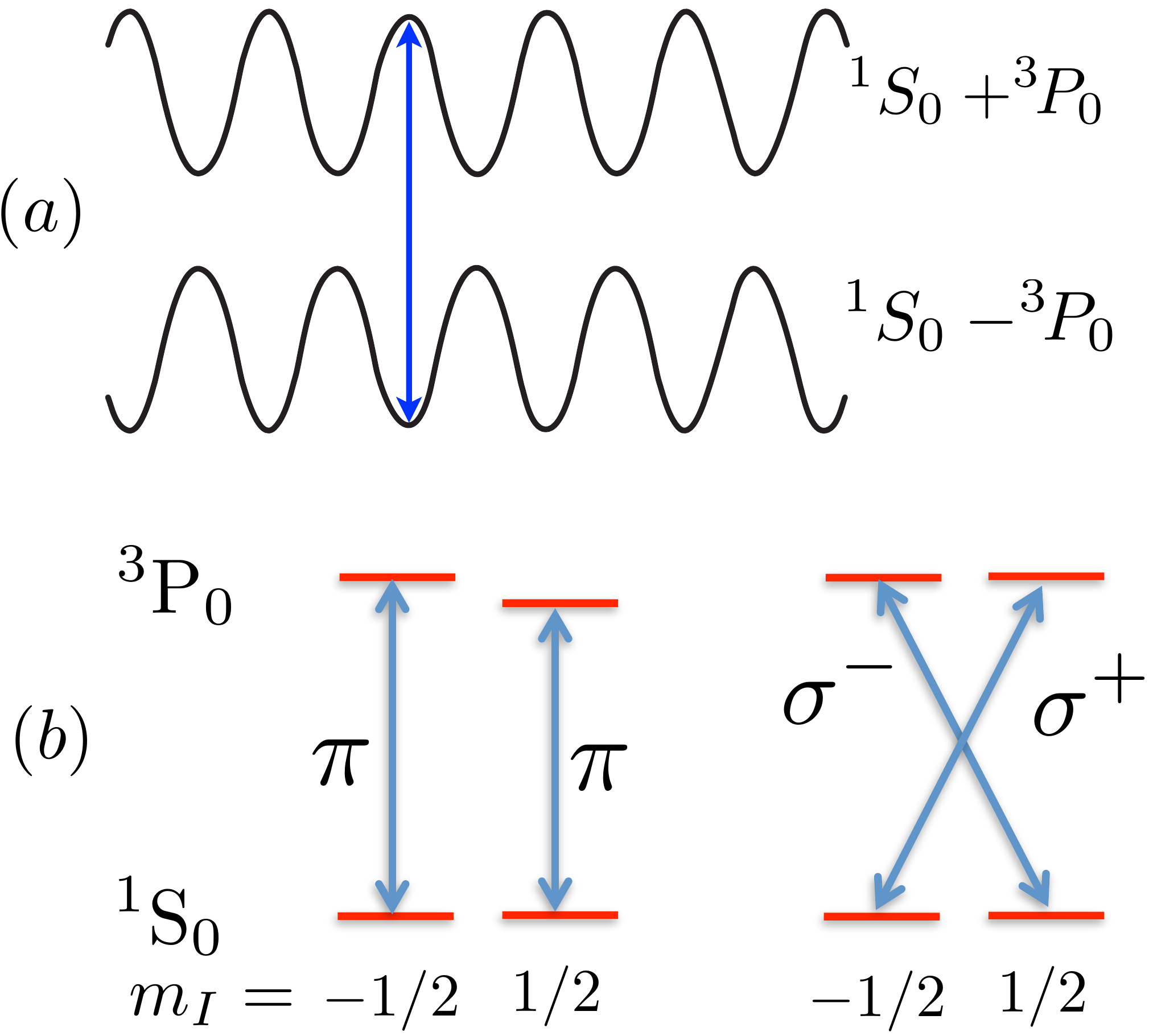}
\caption{Making nuclear-spin-dependent lattices. (a) By coupling on the clock transition with a near-resonant standing wave, it is possible to make dressed potentials for states that are superpositions of $^1$S$_0$ and $^3$P$_0$ levels. (b) These lattices can be made dependent on the nuclear spin either (left) in a strong magnetic field by shifting the transition frequencies \cite{aelong}, or (right) in a weak magnetic field, by using different circularly polarised transitions \cite{yi08}. This second option would be particularly useful in the case of $^{171}$Yb, where $I=1/2$, and these transitions are always closed (i.e., do not couple to other $m_I$ states). } \label{fig:nucspindep}
\end{figure}

Additional schemes can also be created for spin-dependent lattices making use of the metastable $^3$P$_0$ level. In particular, in a strong magnetic field, nuclear-spin dependent lattices can be formed by near-resonant lattices on the metastable transition, creating dressed potentials \cite{aelong} (see Fig.~\ref{fig:nucspindep}a). These can be made nuclear-spin-dependent because the transition frequencies are substantially different for different $m_I$ states in a \emph{strong} external magnetic field (see Fig.~\ref{fig:nucspindep}b, left). Other related schemes have also been proposed in which dressed potentials can be created for $^{171}$Yb using independent lattices with $\sigma^+$ and $\sigma^-$ polarisation components  (see Fig.\ref{fig:nucspindep}b, right). In contrast to the scheme based on frequency shifts, this scheme will work only in the limit of a \emph{weak} external magnetic field.

\subsection{Preparing quantum registers with group-II atoms}

Quantum registers for group-II atoms can be prepared using techniques that have been already developed in the context of group-I atoms. Our focus here will be on proposed schemes involving group-II atoms in optical lattices, with a single atom trapped at every site of a deep optical lattice, initialised in a chosen electronic and nuclear spin state. Fermionic atoms will be the natural choice, because of their non-zero nuclear spin.

A quantum register can be prepared beginning from a degenerate Fermi gas that is optically pumped into a particular nuclear spin level in the ground state $^1$S$_0$ manifold, by adiabatically introducing an optical lattice potential. This potential can be either a magic wavelength lattice \cite{AEatom1}, or a lattice designed for trapping the $^1$S$_0$ manifold only, as discussed in subsection \ref{subsec:control}. The degenerate gas is required so that atom density is sufficiently high to load the lattice with unit filling. A high-fidelity quantum register can then be formed by creating a  band-insulator state \cite{esslinger04}. Using fermionic atoms provides several advantages over loading bosonic atoms, as the energy gap to higher bands in the lattice protects the states from excitations, and in the presence of a harmonic trapping potential, defects in the state are expected to be localised near the edges of the trap \cite{calarco04}. Further improvements to the state could be made, e.g., by filtering \cite{rablloading} or fault-tolerant loading schemes \cite{agloading}.

Optical pumping to produce a pure nuclear spin state was already demonstrated in the context of clock experiments with Sr in Boulder (see, e.g., \cite{ludlow08}), and has recently been demonstrated with degenerate gases of Yb and Sr (see, e.g., \cite{Takahashi, Tey2010dgb}). As mentioned in the introduction, band insulators of Fermions have already been observed in experiments with Yb by the Kyoto group \cite{Takahashi}, and groups at Innsbruck and Rice Universities are planning to introduce this with recently realised degenerate Fermi gases of Sr \cite{DeSalvo2010,Tey2010dgb}. In addition, the high-fidelity arrays of single bosonic atoms on lattice sites that have been clearly demonstrated by various groups using direct imaging \cite{addweiss,addgreiner,addbloch} are very encouraging, and in the fermionic case we expect the fidelity, if anything, to be higher \cite{calarco04}.

\subsection{Quantum Gates}

Single qubit gates can be performed straight-forwardly based on the control developed over electronic and nuclear spin states that was discussed in Sec.~\ref{subsec:control}. Many schemes for two-quibit gates that were proposed originally for group-I atoms can then be improved or freed from technical restrictions by using group-II atoms. For example, Rydberg gates \cite{int:rydberg}, which have been recently demonstrated for trapped group-I atoms \cite{Urban09,Gaetan09} could be directly applied in the context of group-II atoms. The separate hierarchy of Rydberg states for the singlet and triplet manifolds in group-II would also facilitate easier state-dependent excitation in these species, as the electronic-state dependent excitation of either $^1$S$_0$ or $^3$P$_0$ states to a Rydberg level could be performed very cleanly. Gate operations for group-II atoms could similarly be performed using superexchange interactions for fermions \cite{exchangegate}, and in addition Hayes, Julienne, and Deutsch \cite{hayes07} have suggested a gate scheme producing a $\sqrt{{\rm swap}}$ gate based on particle exchange that takes advantage of the nuclear spin degree of freedom. A gate based on particle exchange in which electronic degrees of freedom are manipulated independent of nuclear spin degrees of freedom was proposed by Gorshkov et al. \cite{Gorshkov2009} (see Sec.~\ref{subsec:register} below). Shibata et al. \cite{Shibata} have also proposed the use of magnetic dipole-dipole interactions between two neighbouring atoms excited to the $^3$P$_2$ manifold, which are sufficiently strong for atoms loaded into short period optical lattices, where the lattice spacing $\lesssim 300$ nm.

The spin-dependent lattices discussed in Sec.~\ref{subsec:control} could be applied to implement controlled phase gates via controlled collisions \cite{spindep1}, which has been demonstrated in a proof-of-principle experiment with group-I atoms \cite{spindep2}. Here we would take advantage of 2D or 3D state-dependent lattices in which we would not have to tune trapping lasers between fine-structure states in the way that is required for group-I atoms. 

In both exchange gates and gates via controlled collisions, we require non-zero scattering length for two species. Most fermionic group-II atoms have a sufficiently large scattering length in the $^1$S$_0$ manifold (e.g., $^{87}$Sr) or between atoms in the $^3$P$_0$ and $^1$S$_0$ manifolds. For other species and isotopes with small scattering lengths (e.g., $^{171}$Yb), this could also be achieved using optical Feshbach resonances \cite{srfeshbach,ybfeshbach,entanglefeshbach,pwavefeshbach} to enhance the collisional interaction. An alternative proposal for producing the phase in controlled collisions would be to use blockade gates \cite{aeshort,aelong} that make use of strong interactions or losses when two atoms are present on the same site in the $3$P$_2$ manifold \cite{greene1,greene2}.

In addition, it would be possible to use the electronic state for interfaces in the sense of producing of flying qubits \cite{Gorshkov2009}.

\subsection{Addressing qubits}
\label{subsec:address}
\label{subsec:addressing}

\begin{figure}[tb]
\includegraphics[width=5.5cm]{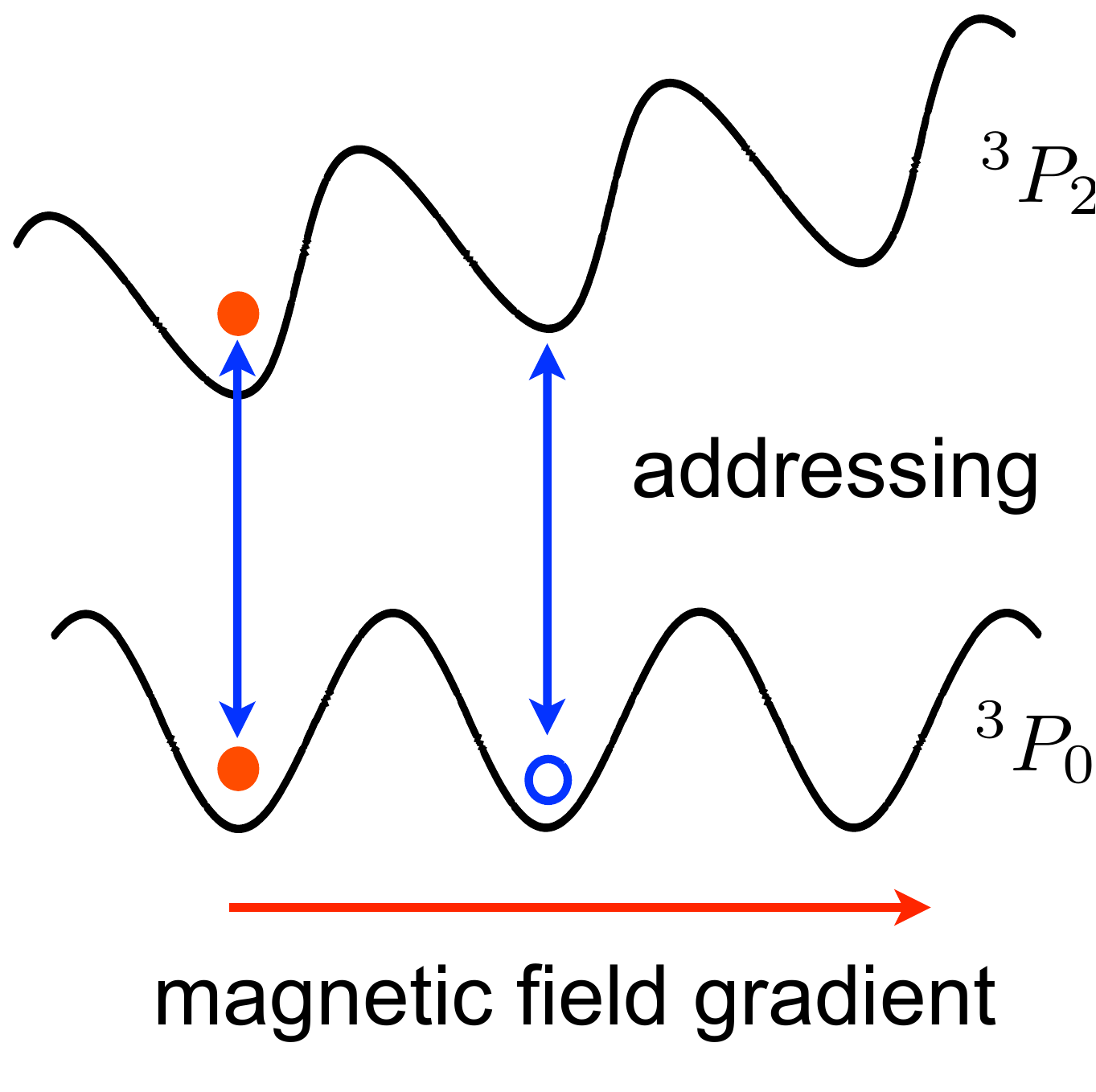}
\caption{Schematic diagram of addressing individual qubits by shifting the energy of states in the $^3$P$_2$ manifold using a magnetic field gradient \cite{dereviankoaddressing}. Coupling of atoms from the $^1$S$_0$ or $^3$P$_0$ manifolds to $^3$P$_2$ (e.g., using a two-photon process) will be off-resonant except at chosen sites (see text for more details). } \label{fig:readoutlevels}
\end{figure}

It is important to have a means to address a single qubit, both for performing gate operations and for reading out the state of the qubit. One possibility would be to directly apply techniques that have recently been developed for addressing in optical lattices with group-I atoms to the group-II species \cite{addbloch2,addweiss,addchin,addgreiner,addgreiner2, addbloch,addott,addmeschede}. However, the metastable levels of group-II atoms can also be used to facilitate addressing in a short-period lattice without the need for the microscope setup of Refs.~\cite{addbloch2, addgreiner2}. Specifically, states in the long-lived $^3$P$_2$ manifold are much more sensitive to magnetic fields than the $^3$P$_0$ and $^1$S$_0$ levels because of their non-zero electron spin. By applying a magnetic gradient field, the energy of these states can thus be shifted relative to the $^3$P$_0$ and $^1$S$_0$ levels in a spatially-dependent manner, making possible a spatially-selective readout or manipulation of qubit states in a form first noted in Ref.~\cite{dereviankoaddressing}. A gradient field of 1 G/cm provides an energy gradient of 4.1 MHz/cm for the $\ket{^3{\rm P}_2,F=13/2,m_f=-13/2}$ level, corresponding to an energy difference of $\sim$15 kHz between atoms in neighbouring sites for a field gradient of 100 Gauss/cm and a $400$ nm lattice. 

An important requirement here is that the states in the $^3$P$_2$ manifold to which the qubit states are transferred is trapped in a lattice at the same position as the $^3$P$_0$ and $^1$S$_0$ levels. If not, an additional trapping laser would be required to compensate this. However, in many cases, this requirement is fulfilled naturally. In Refs.~\cite{aeshort,aelong} it is shown that at the wavelengths used to produce state-dependent lattices for $^{87}$Sr - either independent lattices for $^3$P$_0$ and $^1$S$_0$ or independent lattices for different nuclear spin states by coupling near-resonantly on the clock transition - states in the $^3$P$_2$ manifold are trapped, and no additional trapping for the $^3$P$_2$ manifold is needed for gradient addressing.

Note that as an alternative to using magnetic field gradients for addressing, it would be possible to shift levels by applying a laser with spatially varying intensity at a wavelength that provides a position-dependent differential AC-Stark shift between the $^3$P$_0$ and $^1$S$_0$ levels and the $^3$P$_2$ level \cite{aelong}.

\subsection{A quantum register on a single atom}
\label{subsec:register}

Due to the large number of available nuclear spin states it is possible to encode multiple qubits on one atom. For example, in $^{87}$Sr with $I=9/2$ it is possible to encode up to 3 qubits using 8 of the 10 $m_I$ sublevels in addition to one qubit on the electronic state. A proposal by Gorshkov et al.~\cite{Gorshkov2009} has shown how this system can be treated as a small quantum register encoded on a single atom. This has advantages in that operations on a single atom can normally be performed with high fidelity, which relaxes the quantitative requirement in terms of fidelities for gates between atoms in order to achieve scalable quantum networks. Manipulation of states within a single register can be performed via the arbiratary control we have of nuclear and electronic states on a single atom, as discussed at the beginning of Sec.~\ref{subsec:control}. Readout of the electronic qubit can be performed as described in Sec.~\ref{subsubsec:separateuse}, in a manner that does not affect the nuclear spin states. Gorshkov et al.~\cite{Gorshkov2009} also propose a gate operation between two of these small registers involving just the electronic states. This can be achieved for atoms in neighbouring lattice sites based on a resonant tunnelling scheme, in which atoms are allowed to resonantly tunnel to neighbouring sites only if they are both in a chosen electronic state. This is based on the system having different on-site interaction energy shifts for atoms in different electronic states, and by combining this process with local register manipulation, the gate can be made independent of the nuclear spin state.

\section{Analysis of quantum computing schemes with group-II atoms}
\label {sec:completeanalysis}

In this section we discuss complete quantum computing schemes with group-II atoms. We begin by summarising the general advantages of group-II atoms, illustrated with two examples of complete schemes. We then analyse the key imperfections and sources of decoherence likely to arise in experiments.

\subsection{Discussion of complete quantum computing schemes}

For the purposes of illustration, we now give two possible complete schemes for quantum computing with group-II atoms:

Scheme 1:
\begin{itemize}
\item{Group-II atoms are trapped in a magic wavelength lattice, in a 2D array}
\item{Addressing is performed optically via a microscope}
\item{Gates are performed between atoms using a Rydberg blockade}
\end{itemize}

Scheme 2 \cite{aeshort}:
\begin{itemize}
\item{Group-II atoms are trapped in a 2D array in independent lattices for $^1$S$_0$ and $^3$P$_0$ manifolds}
\item{Addressing is performed via transfer to the $^3$P$_2$ level in a magnetic field gradient}
\item{Gates are performed via controlled collisions, using state-dependent lattices}
\end{itemize}

Both of these schemes benefit strongly from the key features of group-II atoms: In either case, qubits would be encoded on the nuclear spin, and thus substantially less sensitive to decoherence due to magnetic field fluctuations than qubits encoded on electronic states. Also, in both cases, the gate schemes are simplified experimentally compared with the case for group-I atoms: In scheme I, transferring qubits state-selectively to the $^3$P$_0$ level provides a simple route to spin-selective Rydberg excitation, and in scheme II, the different polarisabilities of the $^1$S$_0$ and $^3$P$_0$ means that state-dependent lattices can be formed without lasers tuned near resonance (in the middle of the fine-structure splitting), as is necessary for group-I atoms (see Sec.~\ref{subsec:control} for more details). These are just examples of the possibilities, which were discussed in detail in Sec.~\ref{sec:elements}.

The other key advantage for either of these schemes is that with neutral atoms in optical lattices, it is possible to create a register of many qubits (arrays of thousands of atoms are produced in insulating states on a regular basis in optical lattice experiments, and arrays of the order of 100 - 1000 atoms in 2D have already been demonstrated with direct imaging \cite{addweiss,addgreiner,addbloch}). The difficulty with this top-down approach was always the question of addressing individual atoms, but this can be overcome either via a high-resolution microscope \cite{addbloch2,addgreiner2} or magnetic gradient field addressing (see Sec.~\ref{subsec:address}).

The advantage of Scheme 1 over Scheme 2 is the use of fast Rydberg gates, that could be performed on timescales of microseconds \cite{int:rydberg}. In contrast, the gate times for Scheme 2 are limited as the gates must be much slower than the lattice trapping frequency, which would typically be tens to hundreds of kiloHertz. However, the operations in Scheme 2 could be performed on a massively parallel scale, with every atom involved in every gate operation. In this way, only two operations would need to be performed to create a 2D cluster state, which is a universal resource for measurement-based quantum computation \cite{measurementqc, measurementqc2}. In this sense, Scheme 1 would perhaps be a better candidate for the circuit model of quantum computing, and Scheme 2 for measurement-based quantum computing.

\subsection{Imperfections and sources of decoherence}

In implementing quantum computing with group-II atoms, there will be a number of sources of decoherence that should be controllable in experiments. The key generic sources of heating and decoherence will include magnetic field fluctuations, spontaneous emissions, classical noise on the lattice potential and collisional losses. In addition, specific gate schemes will be associated with noise an decoherence sources depending on their implementation. We comment briefly on the main sources of decoherence below.

\subsubsection{Magnetic field fluctuations}
The nuclear spin qubits considered here have a natural advantage in terms of their insensitivity to fluctuating magnetic fields, and decoherence rates can be reduced by three orders of magnitude compared with qubits encoded on an electron spin, provided the atom is in the $^1$S$_0$ or $^3$P$_0$ manifold. Though combinations of states that are field-insensitive can be used in group-I atoms to reduce decoherence, here we obtain all of the advantages of access to the metastable electronic states, while the natural qubit states, i.e., $m_I$ sublevels are field insensitive and straight-forward to manipulate. For fluctuations with a change in field $\Delta B < 10^{-3}$ G, the differential shift of qubit states encoded in neighbouring nuclear spin levels is $\Delta\omega_{B}<0.3$ Hz for $^{87}$Sr, as the Zeeman shift is $-185$ Hz/G in the $^1$S$_0$ level, and $-295$ Hz/G in the $^3$P$_0$ level. We note that although several schemes for separate manipulation of electronic and nuclear states require large applied fields, the fluctuations of these fields can very often be controlled well in experiments (via measurement and feedback) up to the level of background fluctuations. 

\subsubsection{Spontaneous emissions}
Spontaneous emission events can arise either from decay of an atom excited to one of the metastable electronic manifolds, or due to incoherent scattering of the light generating the lattice. These events can decohere qubits, or cause atoms to be heated to higher motional levels (or even out of the lattice). In the case of decay of a metastable level, the qubit information encoded on the nuclear spin can decohere. As discussed in Sec.~\ref{subsec:control}, the finite lifetime of the $^3$P$_0$ states in fermionic isotopes is generated primarily by hyperfine mixing, which allows flipping of the nuclear spin as atoms decay to the $^1$S$_0$ manifold. However, the lifetime of the metastable states is many seconds, and this source of decoherence can be further suppressed by using these states only for short processes in gate operations. Moreover, in far-detuned lattices, scattering from the lattice light will generally not lead to any change in or decoherence of the nuclear spin state for atoms in the $^1$S$_0$ manifold, in analogy with the electronic state detection scheme discussed in Sec~\ref{subsec:control} \cite{Gorshkov2009}. In this sense, the strongest effect would probably be heating of the motional state of the atoms, but this will be very small in deep optical lattices, as transfer to higher motional states is suppressed by a Lamb-Dicke factor, as in ion traps \cite{ions,ions2, reichenbach07}. Any heating of this kind could also potentially be combatted using the laser cooling scheme proposed in Ref~\cite{reichenbach07}.

\subsubsection{Classical noise on the lattice potential}

Similarly, laser intensity fluctuations and/or jitter of the optical lattice potential can, in principle, give rise to heating of the motional state, as can non-adiabatic processes when transporting atoms in spin-dependent lattices. This could cause problems for gates via controlled collisions, as the interaction energy shift for two particles on the same lattice site will in general depend on their motional state. It could also be an issue for implementation of addressing via magnetic gradient fields, where the addressability depends on energy selectivity of the states. If the lattice potential is not of identical depth for different internal states, it will be important to ensure that all atoms remain in the lowest vibrational level.

However, in a deep lattice, these effects should be very small \cite{heating1,heating2}, and heating of this kind could be combatted using the laser cooling scheme proposed in Ref~\cite{reichenbach07}.

 \subsubsection{Collisional losses from $^3$P$_0$}
Effects of collisional losses from atoms in metastable states have been measured recently in several experiments \cite{lossmeasuresr, bishof2011}. These losses should not play a significant role in decoherence processes here, as two atoms are seldom brought onto the same lattice site. The exception to this would be in gates via controlled collisions. Whilst loss rates from collisions between two atoms in the $^3$P$_0$ manifold can be large, this would never occur in most schemes, as we would normally have at least one atom in the $^1$S$_0$ manifold (with the exception of the scheme in Ref.~\cite{aelong}. For gate times on the order of $1$ms, we would then require collisional stability of atoms on timescales $\sim 100$ ms or longer in order to achieve gate fidelities $\mathcal{F}>99$\%. Measurements of collisional losses between atoms in the $^3$P$_0$ manifold and the $^1$S$_0$ manifold are underway in several groups, in order to determine collisional lifetimes at typical lattice densities ($\sim 10^{14}$ cm$^{-3}$-$10^{15}$ cm$^{-3}$ onsite). 

\section{Analogue Quantum Simulation with Group-II atoms}
\label {sec:simulation}

Group-II atoms have generated special interest in the context of analogue quantum simulation, in which models of interest in many-body physics are engineered in an experiment, and their properties studied via direct experimental measurements. In this section, we briefly summarise some of the new opportunities provided by group-II atoms in this context, especially the possibility to implement high-symmetry models due to the weak dependence of interparticle interactions on the nuclear spin states, and new opportunities for manipulation of atoms in this context provided by the existence of metastable states. 

\subsection{SU(N) Physics with group-II atoms}

In analogue quantum simulation with group-II atoms, the largest interest has focussed on the possibility to create models with SU($N$) symmetry for relatively large $N$. This arises from the fact that at cold temperatures, interactions between atoms are essentially independent of the nuclear spin state. This has been observed in experiments (see, e.g., \cite{kitagawa}), and the possibility to investigate many-body physics in this context was first discussed in detail by Gorshkov et al.~\cite{Gorshkov}, and Cazalilla, Ho, and Ueda \cite{Cazalilla}. 

As a result, fermionic group-II atoms loaded into the lowest band of an optical lattice would be described by a Hubbard model with a Hamiltonian of the form \cite{Gorshkov}
\begin{eqnarray}
H&=&- \sum_{\langle i, j \rangle, \alpha, m}J_\alpha (c^\dag_{i \alpha m}c_{i\alpha m}) + \sum_{j,\alpha} \frac{U_{\alpha \alpha}}{2} n_{j\alpha} (n_{j \alpha} -1) \nonumber \\
& &+ V \sum_j n_{j e} n_{j g} + V_{\rm exch}\sum_{j,m,m'}c^\dag_{jgm} c^\dag_{jem'} c_{jgm'}c_{jem}. \nonumber
\end{eqnarray}
Here, $ c^\dag_{i \alpha m}$ ($c_{i\alpha m}$) is a fermionic mode operator creating (annihilating) a particle on site $i$ in electronic state $\alpha \in \{g,e\}$ and nuclear spin state $m$, the number operator counting particles on site $j$ in the electronic state $\alpha$ is $n_{j\alpha}=\sum_{m} c^\dag_{j\alpha m} c_{j\alpha m}$; $J$ is the tunnelling rate (which is equal for all states as the atoms have the same mass), $U_{\alpha\alpha}$ the interaction between particles in the same electronic state $\alpha$ \footnote{Note that the interaction between two atoms in an excited electronic state $U_{ee}$ should be taken with care, as if these interactions are allowed to occur, they are normally accompanied by large collisional loss.}, and $V$ and $V_{\rm exch}$ the interaction constants for direct and exchange interactions between particles in the electronic states $g$ and $e$. The high degree of symmetry is immediately clear from this model - for $^{87}$Sr, where $I=9/2$, the independence of the Hamiltonian parameters on the nuclear spin state $m$ means that this model can have SU($N$) symmetry up to $N=10$. We note that in an experiment, the degree of symmetry can be controlled (i.e., reduced from the maximum possible value) by varying the number of nuclear spin states that are initially populated. As interactions will not change the nuclear spin, $m_I$ states that are initially unpopulated will remain unpopulated during the experiment. By taking various parameter limits, a variety of spin models can be produced based on this Hamiltonian \cite{Gorshkov}, and these models explored as the SU($N$) symmetry is varied from small to larger values of $N$.

Already before the possibility to use nuclear spins in group-II atoms, there was a lot of theoretical interest in studying systems with spin symmetry in cold gases (see, for example \cite{aecomp0,aecomp0rev}), especially in the context of attractive three-component Fermi gases where the competition between formation of trions and pairing that gives rise to colour-superfluid-like phases has been studied in detail \cite{ae3comp1,ae3comp2,ae3comp3,ae3comp4,ae3comp5,ae3comp6}. Specifically in the context of group-II atoms, Gorshkov et al.~\cite{Gorshkov} proposed the possibility to study a range of SU($N$) spin models, ranging from SU($N$) ferromagnets to the Kondo Lattice model, and Cazalilla, Ho, and Ueda \cite{Cazalilla} investigated the physics of instabilities in a Fermi liquid with high spin symmetry, leading to ferromagnetic states with topological excitations. Hermele, Gurarie and Rey \cite{Hermele2009} have also shown that in a large $N$ limit, chiral spin liquids can be expected from these models, and could potentially be detected in experiments. These developments have led on to a rapid growth in the theoretical literature, opening possibilities for many types of strongly interacting physics involving high degrees of symmetry to be observed in these systems (see, for example \cite{FossFeig,ae3sun,ae3sun1,ae3sun2,ae3sun3,ae3sun4}). This includes not only interesting low-temperature quantum phases, but also non-equilibrium dynamics, such as the possibility to study spin-charge separation in the presence of high degrees of symmetry \cite{ae3sunspincharge}. Progress is also being made in understanding the thermodynamics related to these phases \cite{sunthermodynamics}, and cooling and loading techniques proposed for group-I atoms \cite{coolingreview} should be useful in producing the low temperatures required to see a lot of the most interesting physics in these systems.

\subsection{Other advantages of group-II atoms in quantum simulation}

Metastable levels of group-II atoms also provide an important tool for analogue quantum simulation, and can be used as auxilliary levels in a number of contexts. For example, these states can be taken advantage of in schemes to form artificial gauge fields \cite{Gerbier2010}, or to produce sub-wavelength optical lattices \cite{yi08}. The use of the metastable state has also been proposed for the introduction of a controlled dissipative mechanism \cite{DiehlYi2010}, in which atoms selectively undergo spontaneous emission events in a manner that does not change their nuclear spin state (see Sec.~\ref{subsec:control}). This mechanism can be used as part of a scheme to drive atoms in an optical lattice into desired many-body states, in the case of Ref.~\cite{DiehlYi2010}, a state exhibiting d-wave pairing between particles of different nuclear spin.

\section{Summary and Outlook}
\label{sec:summary}
In summary, group-II atoms offer many new possibilities to address current challenges in quantum computing and quantum simulation with neutral atoms. While experiments with these species are relatively recent, very precise control over the electronic and nuclear spin states has been demonstrated in the context of optical clocks, and already insulating states of Yb have been produced in optical lattices. This opens the possibility for a new generation of quantum computing experiments with neutral atoms, taking advantage of the nuclear spin degree of freedom in order to store qubits while minimising the decoherence due to external field fluctuations, and making use of the independent electronic degree of freedom on the clock transition for manipulation and readout of qubits. Many gate schemes proposed in the context of group-I atoms can be freed from technical restrictions by using group-II atoms, and new possibilities are opened, e.g., the laser cooling of atoms using the electronic state, whilst preserving the qubit state stored on the nuclear spin. At the same time, developments in this field will also be driven by interest in analogue quantum simulation, with exciting prospects for the realisation of strong-interacting many-body models with a high degree of symmetry, due to the independence of interactions on the nuclear spin in these systems. 

\begin{acknowledgements}
I would like to thank Alexey Gorshkov, Ana Maria Rey, Jun Ye, and Peter Zoller for discussions and collaborations on quantum computing and quantum simulation with group-II atoms over the last few years, which strongly influenced this review. 
\end{acknowledgements}

\end{document}